\documentclass[conference]{IEEEtai}
\IEEEoverridecommandlockouts
\usepackage{cite}
\usepackage{amsmath,amssymb,amsfonts}
\usepackage{algorithm}
\usepackage{algpseudocode}
\usepackage{graphicx}
\usepackage{tabularx}
\usepackage{textcomp}
\usepackage{xcolor}
\usepackage{booktabs}
\usepackage{ifthen}
\usepackage{tikz}
\usepackage{soul}
\usetikzlibrary{matrix,calc}
\usepackage{url} 

\def\BibTeX{{\rm B\kern-.05em{\sc i\kern-.025em b}\kern-.08em
    T\kern-.1667em\lower.7ex\hbox{E}\kern-.125emX}}

\begin{document}

\title{
Extensible Machine Learning for Encrypted Network Traffic Application Labeling via Uncertainty Quantification
}

\author{\IEEEauthorblockN{Anonymous Author(s)} \\
\IEEEauthorblockA{Anonymous Address\\ Email: author@email}}
\author{\IEEEauthorblockN{Steven Jorgensen, John Holodnak, Jensen Dempsey, Karla de Souza, Ananditha Raghunath, \\ Vernon Rivet, Noah DeMoes, Andr\'es Alejos, and Allan Wollaber\IEEEauthorrefmark{1}}\IEEEauthorblockA{MIT Lincoln Laboratory, Lexington, Massachusetts\\ Email: allan.wollaber@ll.mit.edu\IEEEauthorrefmark{1}}}

\maketitle
\thispagestyle{plain} 
\pagestyle{plain} 

\begin{abstract}
With the increasing prevalence of encrypted network traffic, cyber security analysts have been turning to machine learning (ML) techniques to elucidate the traffic on their networks. However, ML models can become stale as new traffic emerges that is outside of the distribution of the training set. In order to reliably adapt in this dynamic environment, ML models must additionally provide contextualized uncertainty quantification to their predictions, which has received little attention in the cyber security domain. Uncertainty quantification is necessary both to signal when the model is uncertain about which class to choose in its label assignment and when the traffic is not likely to belong to any pre-trained classes.

We present a new, public dataset of network traffic that includes labeled, Virtual Private Network (VPN)-encrypted network traffic generated by 10 applications and corresponding to 5 application categories. We also present an ML framework that is designed to rapidly train with modest data requirements and provide both calibrated, predictive probabilities as well as an interpretable ``out-of-distribution'' (OOD) score to flag novel traffic samples. We describe calibrating OOD scores using p-values of the relative Mahalanobis distance.

We demonstrate that our framework achieves an F1 score of 0.98 on our dataset and that it can extend to an enterprise network by testing the model: (1) on data from similar applications, (2) on dissimilar application traffic from an existing category, and (3) on application traffic from a new category. The model correctly flags uncertain traffic and, upon retraining, accurately incorporates the new data.

\end{abstract}

\begin{IEEEkeywords}
network traffic classification, virtual private networks, machine learning, discrete wavelet transform, encrypted traffic, cybersecurity, uncertainty quantification
\end{IEEEkeywords}

\section{Impact Statement}
Labeling application-generated, VPN-encrypted network traffic is challenging given limited public datasets, the difficulty of signal-extraction using only packet size and timing information, and the need to account for uncertainty in the machine learning (ML) pipeline to screen out predictions for novel applications or low-confidence predictions. We provide a new public, labeled packet capture dataset with VPN-encrypted network traffic for researchers to train their own models, filling a needed gap for this problem domain. We propose data features and a neural network architecture that accurately trains with modest data requirements and provides predictions alongside confidence scores that allow analysts to know when a model is unsure about a given prediction. By integrating data, feature generation, and confidence-aware ML, we have laid a foundation that will allow future researchers to create robust analytics for this challenging domain at the intersection of cyber security and ML.

\section{Introduction}
The proportion and popularity of encrypted network traffic has quickly risen over the last several years, with over 95\% of traffic across Google and 97 of the top 100 websites defaulting to encryption \cite{GoogleTransparency}. This is a boon for online privacy; however, there is a corresponding increase in encryption for malware delivery \cite{gartner2017security, Watchguard21}. Beyond that, MITRE ATT\&CK enumerates several techniques in which encryption has been used to obfuscate command and control or to masquerade one application as another. Although signature-based techniques such as website fingerprinting \cite{rimmer2017website, hoang_domain_2021} and (e.g.) Palo Alto's App-ID can flag particular packet sequences that indicate a known website request or protocol handshake, the general problem of inferring an application when signatures and heuristics fail remains a challenging problem. In particular, Virtual Private Network (VPN) providers have also lowered the barriers for users not only to encrypt the ports, protocols, and IP addresses of their connections, but also to obfuscate the IP address of the VPN server itself \cite{NordVPN} and/or pad all packets to be the same size before encryption \cite{NoiseProtocol}, which can defeat signature detection capabilities. In light of these difficulties, network security analysts have begun turning to machine learning (ML) approaches that leverage latent signals in the packet timings, sizes, and their encrypted payloads in order to predict the applications that generated the encrypted traffic. 

Until recently, existing work in machine learning for encrypted traffic classification has focused primarily on feature construction and model development to optimize raw performance metrics such as overall accuracy.  
For example, traditional machine learning models (like na{\"i}ve Bayes and decision trees) have been applied to simple statistical features \cite{alshammari2009machine, draper2016characterization} and neural networks based on one or two-dimensional convolutions have been applied to sequences of raw byte values, interarrival times, and wavelet coefficients \cite{liang2019content, Abbasi2021, Wang2017a, Wang2017b, Rezaei2019, Lotfollahi2020, Akbari2021}.  Many of these techniques yield very good performance on publicly available data, the most popular being the ISCXVPN2016 dataset for studying VPN encryption \cite{draper2016characterization}. Much more recently, end-to-end and transformer-based architectures have begun showing impressive performance \cite{liu_fs-net_2019,aceto_mimetic_2019,wang_multi-scale_2022,aceto_distiller_2021} on a variety of IoT and mobile traffic datasets outside of a VPN setting
\cite{aceto_mirage_2019,heng_utmobilenettraffic2021_2021,zaki_grano-gt_2021}.

In this work, we build upon some previously developed features (interarrival statistics and wavelet signal processing) with an eye towards enabling rapid learning with limited data. In addition, we focus on developing an ML model with accurate uncertainty quantification that is able to (1) produce accurate probabilistic predictions for the classes on which the model is trained, and (2) detect examples dissimilar to application categories that the model was trained to predict.
While there has been some limited mention of uncertainty quantification in cyber security applications of machine learning in general \cite{Highnam2021, Brown2020, Engel2015, Darling2018, Nascita2021}, the problem has not received the same attention in cyber security as it has in other domains, such as computer vision and healthcare.  We consider this unfortunate and believe that in a dynamic environment such as a computer network it is critically important to quantify model uncertainty.
 A recent review highlighted the need to reduce errors (false positives) when moving from a closed training set to real-world data for encrypted network traffic analysis \cite{Review2021}, and Nascita et al.'s, emphasis on trustworthiness and explainability in deep learning in traffic classification is a step in the right direction \cite{Nascita2021}.

In our work, we develop a classifier for encrypted network traffic based on prototypical networks, which leverage distances to class prototypes (means) in a learned embedding space to compute class probabilities and thus predictions \cite{snell2017prototypical}.  We use the recently developed relative Mahalanobis distance between test examples and class prototypes to detect out of distribution (OOD) examples \cite{Ren2021}.  Our model uses as input a mixture of simple features derived from flow statistics as well as wavelet transform coefficients to produce class predictions for segments of bi-directional network flows (although more advanced, transformer architectures such as in \cite{wang_multi-scale_2022} should also be compatible with a prototypical network layer). To enable regular predictions in VPN settings in which (1) the embedded 5-tuples for connections are encrypted and (2) the beginnings and endings of connections are obfuscated, we split all network flows into discretely sampled time segments and make one prediction for each time segment in each flow.  

We apply our model to a new dataset (which we have made publicly available) of encrypted network traffic collected on a testbed from ten applications that cover five broad, but not comprehensive categories of traffic.  This dataset supplements those already available in the literature, which either cover only very specific types of network traffic such as the UC Davis dataset of QUIC traffic \cite{Rezaei2019} or the Orange dataset of HTTPS traffic \cite{Akbari2021}, or contain artifacts such as unencrypted packets that could affect the validity of models trained on them\footnote{See Section \ref{sec:CIC} for specific examples of artifacts we identified in the often-used ISCXVPN2016 dataset, introduced in \cite{draper2016characterization}, which we hope to supplement.}. When trained with an 80/20 train/test split, our model has a micro F1-score of $0.98$. Further, we investigate our model's performance when training with only small fractions of the dataset, achieving (for instance) 95\% accuracy on VoIP traffic in a 5-class problem using just under 14 minutes of data capture. On the UTMobileNetTraffic2021 dataset, our model has $80$\% accuracy, comparable to Heng et al.'s best results \cite{heng_utmobilenettraffic2021_2021}. 

We stress-test the model by attempting to transfer its learning from the testbed to an enterprise network over several categories. We found that in some instances (a streaming and VoIP application), the model cleanly transfers onto another network with high accuracy and low uncertainty, and in others (two file transfer applications), the model successfully reports low in-distribution confidence and can be quickly retrained to incorporate the new applications for identification in the known category.

To test our model's ability to identify OOD examples, we perform inference on PCAP containing an unseen traffic category (Zoom) from an enterprise network and demonstrate that the model reliably assigns the new category a high OOD score, with over 77\% predicted as significantly out of distribution. Upon retraining with about two hours of labeled examples of Zoom, which takes under 5 minutes, the fraction of significantly OOD test examples drops to 26\%.

Finally, we test our model's robustness by investigating the case where all packets are padded to have identical sizes, as occurs in the Noise Protocol or in certain implementations of Traffic Flow Confidentiality \cite{NoiseProtocol, RFC4303}.  We see that the model (which utilizes both packet timing and size information) is still able to obtain very good performance (micro F1-score of 0.97) indicating that packet size information is not necessary for accurate classification.  In addition, this observation calls into question the degree of privacy offered by such obfuscation strategies.

To summarize, the primary contributions of this paper are:
\begin{itemize}
\item a new public dataset of application-labeled, VPN-encrypted and non-VPN-encrypted network traffic,
\item an ML architecture for encrypted traffic designed to quickly train with limited labeled data and provide calibrated, uncertainty-contextualized predictions for two tasks: (1) predictive uncertainty, when the model is indecisive about which \emph{known} label a sample belongs to and (2) model uncertainty, when the model encounters OOD data potentially from an \emph{unknown} label, enabling the model to be systematically extended to new data, and 
\item a demonstration of sufficiently high accuracy and F1 score for application data that is grouped into discrete time windows, instead of by collecting statistics or data for an entire connection or relying upon fingerprints or signatures, enabling sequential application predictions under a VPN.
\end{itemize}

The remainder of our paper is organized with related work in Sec.\ \ref{sec:rw}, our PCAP dataset in Sec.\ \ref{sec:Dataset}, our data and learning techniques in Sec.\ \ref{sec:methods}, our computational experiments in Sec.\ \ref{sec:results}, a discussion in Sec.\ \ref{sec:discussion}, and conclusions in Sec.\ \ref{sec:conclusions}.

\section{Related Work}\label{sec:rw}
In this section, we describe existing work in the literature related to traffic classification in general, encrypted traffic classification, and website fingerprinting.  Due to our focus on uncertainty in encrypted traffic classification, we briefly review relevant work on uncertainty in deep learning from the machine learning community.  Finally, we discuss existing encrypted traffic datasets and our motivation for collecting our own.

\subsection{Early Traffic Classification}
In the early days of the Internet, censors used port-based traffic classification to restrict Internet content \cite{tschantz2016sok}. Port-based classification models could classify network traffic using the port numbers encoded in the packet headers because the Internet Assigned Numbers Authority maps services to unique port numbers \cite{al2015network}. In response, clever Internet users and application developers adapted to port-based censorship by dynamically changing port numbers, hiding behind port numbers already assigned to well-known, trusted applications, or using unregistered port numbers \cite{callado2009survey, roughan2004class, moore2005ports}. Deep 
packet inspection (DPI) then emerged as a technique that inspects the payloads of packets and classifies them accordingly 
\cite{deri2014, tschantz2016sok, moore2005ports, sen2004payload, choi2004payload}. DPI is often used to check for malicious code, obtain situational awareness on a network, and eavesdrop on connections.
For example, the operators of the Great Firewall of China have used DPI to inform active probing to identify and censor Tor \cite{tschantz2016sok}. However, with the rise of easy-to-use encryption schemes such as HTTPS and VPNs, 
DPI has lost its effectiveness. Although possible, traffic decryption is not a viable countermeasure, because decryption not only compromises user privacy but is also computationally expensive and difficult to implement \cite{anderson2016identifying}. VPNs in particular can effectively obscure payloads, ports, and IP addresses, through encryption, leaving only packet timing and size information as potential signal sources.

\subsection{Encrypted Traffic Classification}
To overcome the limitations of port- and payload-based classification models, researchers have focused on traffic classification models that use features from observable encrypted traffic metadata \cite{anderson2016identifying, Review2021}.

Some examples of features constructed from observable metadata include features based on simple statistics derived from flows \cite{mcgregor2004flow, roughan2004class, zander2005automated, moore2005ports, li2009efficient, yuan2010svm, soysal2010machine, zhang2012feature, fahad2014optimal, draper2016characterization, anderson2016identifying, yamansavascilar2017application, lashkari2017characterization} and wavelet-based features \cite{shi2017wavelet, liang2019content}. While traffic classification methods that use flow statistic-based features yield respectable results for isolated application classification in contained environments, it is uncertain whether they are robust enough to be utilized in real-world settings. Flow statistic-based features can vary widely between network topologies and are not able to characterize the highly variable and bursty nature of Internet traffic \cite{gilbert1999scaling, molnar1999source, riedi1999multifractal, riedi2000toward}. On the other hand, wavelet-based features can capture the inherent nonlinearities of Internet traffic, such as jumps and edges, by providing representations of the observable metadata from the bidirectional connection that are localized in both time and frequency \cite{hastie2009elements}. Shi et al.\ \cite{shi2017wavelet} are the first to use wavelet-based features for traffic classification and show that SVMs trained with wavelet features outperform those trained with flow statistic-based features. Later \cite{liang2019content} used Convolutional Neural Networks (CNNs) trained on wavelet-based features for traffic classification. We build on this approach by dividing the connections into chunks of time rather than classifying the aggregate connection all at once and combining flow statistic-based features and wavelet features in each time window. Roy et al.\ also use a time-based approach via a neural ordinary differential equation on packet interarrival times, and they also avoid payload inspection, which is similar in spirit to our approach \cite{Roy_Fast_2022}. Shapira and Shavitt use a clever time/size technique to generate images (``FlowPic'') from application traffic, which then enables standard image classification techniques  \cite{Shapira_FlowPic_2021}. Recently, end-to-end and and transformer-based architectures have bypassed the need for feature generation, shifting the burden into the neural network, which typically leverages on the order of a million parameters and achieve state-of-the-art accuracies \cite{liu_fs-net_2019,aceto_mimetic_2019,wang_multi-scale_2022,aceto_distiller_2021}. In our work, we achieve high accuracy using fewer than $30,000$ parameters, which enables rapid training (minutes).

\subsection{Website Fingerprinting}
Website fingerprinting is a process of recognizing specific web traffic (e.g., to particular websites) based on unique patterns in the traffic \cite{bhat_var-cnn_2019}. This approach leverages the fact that websites tend to have unique packet sequence patterns (handshakes) that allow for identification. Traffic features can include information such as unique packet lengths, sequence lengths, packet ordering, and packet interarrival timings \cite{rahman_tik-tok_2020}. These features are then fed to ML algorithms that classify traffic based on the observed patterns. However, the identifying handshake must be learned beforehand and observed in a traffic sample to enable detection. Website fingerprinting techniques also tend to be sensitive to changes in network patterns such as packet padding and fluctuations in packet round trip time, and it is challenging to know when the models must be updated \cite{juarez2014critical}. 

We envision our approach would work in tandem with website fingerprinting and signature-based approaches, allowing for contextual, uncertainty-quantified predictions to be made before or after a handshake has been identified.
An example of this is the case where, within some packet capture, website fingerprinting identifies a handshake to a malicious website. Our approach could then also be applied to the connection to determine if any file transfer or new command and control channels occurred after the signature was detected.

\subsection{Uncertainty in Deep Learning}

\begin{table*}[!t]
\centering
\caption{Distance-based approaches to OOD detection.}
\label{tab:OOD}
\begin{tabularx}{\textwidth}{XX}
\toprule
\bf{Approach} & \bf{Comments} \\
\midrule
Mahalanobis distance to class prototypes \cite{Lee2018} \cite{Wang2020} \cite{Maciejewski2022} \cite{Ming2022} \cite{Ren2021} & \hspace{-0.4em} \cite{Ming2022} modifies the loss to separate class prototypes using angular distance \newline \cite{Ren2021} normalizes the distance to the class prototypes using the distance to the overall data distribution  \\
\midrule
Euclidean distance to fixed point in feature space \cite{Huang2021} & Fixed point is the embedding of random noise \\
\midrule
Euclidean distance to nearest neighbors \cite{Sun2022} & Avoids making parametric assumptions (multivariate normality) on the embeddings \\
\midrule
Cosine similarity to weight vectors in last layer \cite{Tech2020} & Modifies loss to use softmax of similarities  \\
\bottomrule
\end{tabularx}
\end{table*}
Quantifying the uncertainty of model predictions is an active area of research in the machine learning community.  Considerable research has been focused on improving the ability of classification models to output calibrated probabilistic predictions and to detect OOD examples at test time.

To be clear, a model's predictive probabilities are calibrated if their confidence (highest predicted probability) matches their accuracy.  In other words, a model should correctly label 90\% of the examples that it categorizes with 90\% confidence.  Guo et al. \cite{Guo2017} showed that while shallow neural networks tend to be reasonably well calibrated, deep neural networks are not.  Various procedures for improving calibration have been proposed including temperature calibration \cite{Guo2017}, deep ensembles \cite{Lakshminarayanan2017}, and Bayesian Model Averaging \cite{Wilson2020}.  In the context of traffic classification, Nascita et al. \cite{Nascita2021} evaluate the calibration of several classifiers in terms of expected calibration error (bin-wise average difference between confidence and accuracy) and class-wise expected calibration error, and their emphasis on trustworthiness and explainability makes their paper nearest in spirit to this work, although it does address OOD detection.

We now briefly review some of the major approaches to OOD detection in the machine learning literature.  The degree to which examples are out of distribution is usually quantified either using the distribution of predicted probabilities for examples or some notion of distance to the distribution of training data.  A detailed review of OOD detection approaches is given in \cite{Yang2022}.

For example, the magnitude of the largest predicted probability was used in \cite{Hendrycks2017} as a score, encoding the assumption that if a data example is out of distribution, no class will be assigned a high probability.  Other work \cite{Liang2018}, improved this approach by applying input preprocessing and softmax temperature scaling to improve separation between the largest predicted probabilities of in and out of distribution examples.  In another approach \cite{Malinin2018}, the authors model the output probabilities as a distribution and place a Dirichlet prior over them.  During training, they attempt to enforce the assumption that uncertain examples produce a flat distribution of predictive probabilities by training on OOD examples.  This idea is improved in \cite{Tsiligkaridis2021}, which removes the necessity for using OOD examples in training.  

On the other hand, the authors in \cite{Lee2018} fit class-conditional Gaussian distributions in the embedding space of a model with softmax output probabilities and compute Mahalanobis distances to the Gaussian distributions at test time.  A very similar approach is taken in \cite{Wang2020}, but in the context of meta-learning and prototypical networks.  In this case, the authors use the distance to the closest class prototype as an OOD score.  Maciejewski et al. \cite{Maciejewski2022} examine the effect of covariance types and the dimension of the embedding space on OOD detection performance.  To encourage separation between classes, the authors in \cite{Ming2022} attempt to minimize the angular distance between a class prototype and examples of that class and maximize the angular distance different class prototypes.  They then use the Mahalanobis distance to the closest class to uncover out of distribution examples.  Other interesting approaches include basing OOD detection on the distance to a fixed point in feature space, representing the embedding of random noise \cite{Huang2021} and computing the cosine similarity between the embedding and the last layer weight vectors \cite{Tech2020}.  Finally, to avoid making distributional assumptions, the authors in \cite{Sun2022} compute the distance to the $k^{th}$ nearest neighbor.

Our approach is also based on the distance of test examples to the training data.  It is most similar to that in \cite{Wang2020}, but instead makes use of the relative Mahalanobis distance from \cite{Ren2021}.  We summarize the distance-based approaches, as they are most similar to ours, in Table \ref{tab:OOD}.

\subsection{Publicly Available VPN-Encrypted Traffic Datasets}
\label{sec:CIC}
The development of reproducible research in VPN-encrypted traffic classification is hindered by the shortage of publicly available VPN-encrypted traffic data. In 2016, the Canadian Institute for Cybersecurity published the ISCXVPN2016 dataset of VPN and non-VPN traffic from 20 applications and seven traffic categories \cite{draper2016characterization}. As the first publicly available dataset of VPN traffic, the ISCXVPN2016 dataset has been used by many researchers to train and test ML models. Other datasets with encrypted traffic have since been released as well, such as the UC Davis dataset of QUIC traffic \cite{Rezaei2019}, the Orange dataset of HTTPS traffic \cite{Akbari2021}, as well as multiple IoT and mobile datasets \cite{aceto_mirage_2019,heng_utmobilenettraffic2021_2021,zaki_grano-gt_2021}, although none of these distinguishes between VPN and non-VPN encrypted data.

Despite its widespread use, we found that the ISCXVPN2016 dataset contains discrepancies in the VPN captures that compromise the integrity of encrypted network traffic classification methods evaluated on it (these are in addition to Aceto et al.'s observation that 65\% of the biflows are due to BlueStacks and should be filtered out \cite{aceto_distiller_2021}). Upon closer inspection of the PCAP files from the VPN captures, we find packets with unencrypted payloads. For example, the payload of the 17th packet in the PCAP file for the ICQ Chat VPN capture in \texttt{vpn\_icq\_chart1b.pcap} contains body HTML text saying ``how are you today'', which can be easily viewed in WireShark. We also find PCAP files for VPN-labeled captures to contain multiple connections, although we expect the PCAP files for a VPN session to contain a single connection between the VPN client and the VPN server. 
As an example, the vast majority of packets in the \texttt{vpn\_netflix\_A.pcap} file are over port 80 to a Netflix IP address, the majority of packets in \texttt{vpn\_hangouts\_audio1.pcap} resolve to a Google IP address, and the majority of packets in \texttt{vpn\_facebook\_audio2.pcap} resolve to a Facebook IP address (with a minority fraction also going to Google). These findings imply that either the network was tapped before the packets went through the VPN client or the packets were not encrypted.

Unencrypted packet payloads will unfairly strengthen traffic classification methods that leverage packet payloads, such as DPI or the deep packet approach in \cite{Lotfollahi2020}. Additionally, the presence of multiple connections per VPN capture will unfairly strengthen methods that leverage connection information aggregated at the flow-level. 

In light of these concerns, we recommend that researchers seeking to utilize the ISCXVPN2016 dataset ensure that their traffic classification methods do not presume that VPN-labeled packet payloads are VPN-encrypted. Additionally, if their methods for analyzing VPN-encrypted traffic rely on connection information aggregated at the flow-level, we recommend that they post-process the packet headers to ensure that all packets appear under the same 5-tuple, as they would inside a VPN connection. Alternatively, one could also replay the VPN-labeled files through a VPN.  

Finally, there is also a challenge with respect to consistent assignment of the traffic categories to the collected PCAP files. As an example, the file {\tt skype\_video1a.pcap} could reasonably be categorized as VoIP or Web Browsing, but probably would have fit better with other video-teleconferencing PCAPs in its own VTC category. 

In response to the limitations of the ISCXVPN2016 dataset, we introduce a supplemental, labeled dataset of VPN and non-VPN network traffic for public use and to test our extensible machine learning framework. 

\section{Encrypted Traffic Dataset}
\label{sec:Dataset}
In this section, we describe the characteristics of our dataset\footnotemark \footnotetext{The dataset is available for download at: \textit{https://www.ll.mit.edu/r-d/datasets/vpnnonvpn-network-application-traffic-dataset-vnat}.} and outline the process used to produce the dataset. Our dataset is a collection of 36.1 GB of labeled network traffic that contains both VPN-encrypted and unencrypted flows from the five traffic categories shown in Table \ref{tab:vast_apps}, which is inspired by the categories in \cite{draper2016characterization} with a few omissions (e.g., P2P, Web Browsing) and a new category (Command \& Control). See Table \ref{tab:vast_files} in Appendix \ref{appendix:filemappings} for the file mappings used to assign class labels.

\begin{table}[htbp]
\caption{The applications and traffic categories used to generate the dataset.}
\label{tab:vast_apps}
\centering
    {
    \begin{tabular}{ll}
    \toprule
    \bf{Traffic Category} & \bf{Applications} \\ \midrule
    Streaming & Vimeo, Netflix, YouTube \\  
    VoIP & Zoiper  \\  
    Chat & Skype  \\  
    Command \& Control & SSH, RDP \\  
    File Transfer & SFTP, RSYNC, SCP \\ \bottomrule 
    \end{tabular}
    }
\end{table}

\subsection{Dataset Characteristics}
Our dataset consists of 36.1 GB, 33,711 connections, and approximately 272 hours of packet capture from five traffic categories as listed in Table \ref{tab:vast_stats}. Although the File Transfer category constitutes 90\% of the total size of the dataset, it only contributes to roughly 40\% and 5\% of the total connections and time, respectively. In contrast, the Command \& Control category only makes up about 2\% of the dataset in size while comprising 40\% of the connections and almost 47\% of the total time. The Streaming category contributes approximately 7\%, 5\%, and 2\% of the total size, connections, and time. The Chat category contributes less than 1\% of the total size, 4\% of the total connections, and 45\% of the total time. The smallest traffic category, VoIP, makes up around 1\% of the total size and time, and less than 2\% of the total connections. Although each protocol is encrypted by default, we also report the percentage of each category that is VPN-encrypted, which varies from 41 to 65\% by size.
In addition to the complete dataset in PCAP form, researchers can also download a pandas dataframe in HDF5 format that contains the connection (biflow) 5-tuple, packet timestamps, directions, and corresponding label (filename containing application and VPN status). 

\begin{table}[htbp]
    \caption{Total packet sizes, connections, and time per traffic category in the dataset, as well as the percentage of each category that is VPN-encrypted.}
    \label{tab:vast_stats}
    \centering
    {
    \begin{tabular}{lcccc}
    \toprule
    \bf{Traffic Category} & \bf{Size (MB)} & \bf{Connections} & \bf{Time (h)} & \bf{VPN \%}\\ \midrule
    File Transfer & 32,601 & 16,430 &  14.8 & 43\\  
    Streaming & 2626 & 1,764 &  4.3 & 41\\  
    VoIP & 103 & 617 & 2.7 & 65 \\  
    Command \& Control & 622 & 13,599 &  127.4 & 49\\  
    Chat & 185 & 1,301 &  123.0 & 63 \\  \midrule
    \bf{Total} & 36,137 & 33,711 & 272.3 & 43\\  
    \bottomrule
    \end{tabular}
    }
\end{table}

\subsection{Network Setup} 
\begin{figure}[!b]
        \centering
        {\includegraphics[width=3.3in]{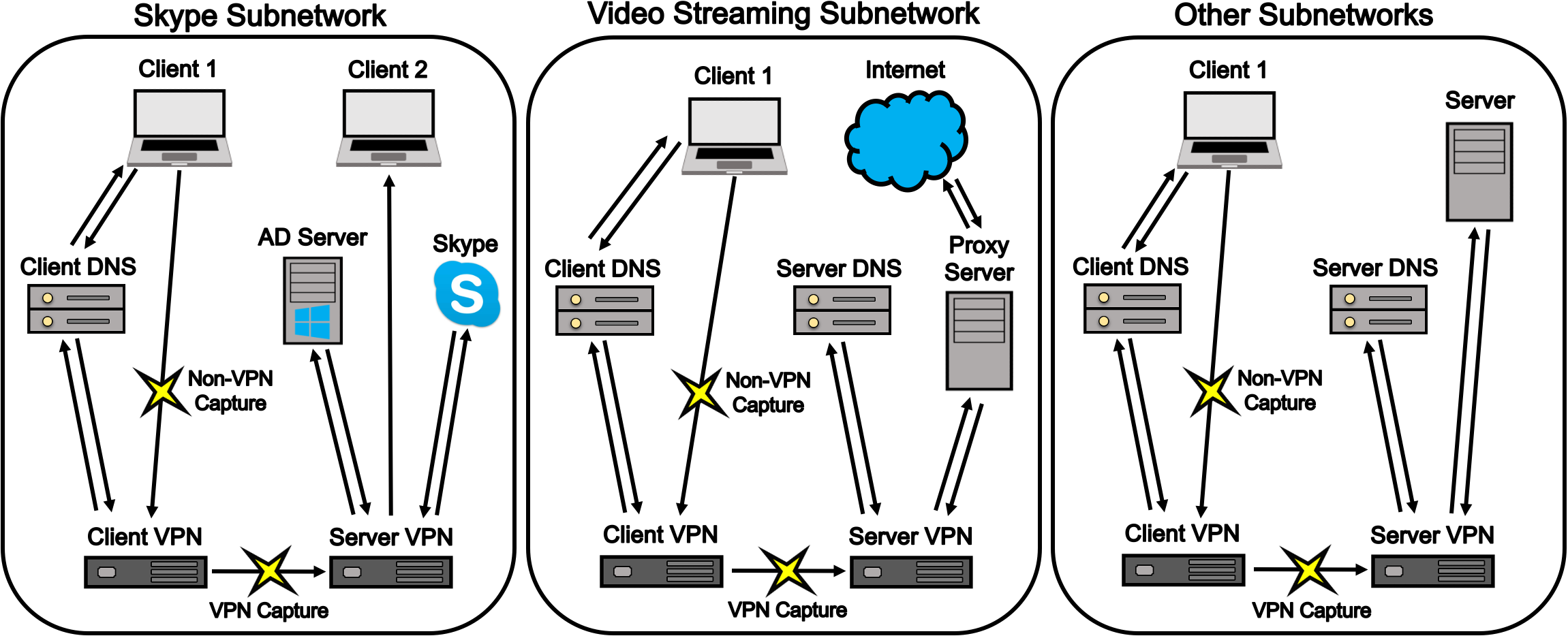}}
        \caption[Diagram of Virtual Subnetworks]{Diagram of the virtual subnetworks used to capture network traffic. The yellow stars denote where the data captures occur. The non-VPN traffic is captured between the client and the client VPN while the VPN traffic is captured between the client VPN and the VPN server.}
        \label{fig:subnetworks}
\end{figure}

To produce the dataset, we begin by creating virtual subnetworks for each traffic category as shown in Figure \ref{fig:subnetworks}. Each subnetwork contains a client, a client DNS server, a VPN client, and a VPN server. The Skype subnetwork contains an additional client to allow for bidirectional chat. The video streaming and web browsing subnetworks are connected to the Internet to enable access to Firefox, Chrome, YouTube, Netflix, and Vimeo. As shown in Figure \ref{fig:subnetworks}, VPN traffic is captured between the VPN client and the VPN server. Separately, non-VPN traffic is captured between the VPN client and the application layer.
    
\subsection{Data Generation}
Once the network setup is complete, we generate the dataset for each traffic category individually to ensure data purity. 
All traffic is captured using tcpdump and stored in the PCAP format. Filename-to-category mappings are provided in Table \ref{tab:vast_files} in the Appendix.

The Streaming category includes Vimeo, Netflix, and YouTube traffic generated by team members watching videos on each platform. The VoIP category is composed of VoIP audio generated by team members having conversations with one another using Zoiper. The Chat category consists of Skype Chat traffic generated with bots that replay actual chats from publicly available logs from Gitter.im chat rooms \cite{chatbot_scripts}. 

The Command \& Control category includes SSH and RDP traffic. RDP traffic is generated by team members actively performing tasks, such as editing Word documents, over an RDP connection. To emulate SSH traffic, we created a script to randomly sample 2 to 20 commands with replacement from a predetermined list of 17 commands. For each command, it samples a random wait time between 1 and 60 seconds. Once it is finished executing each command and wait time, it waits 60 seconds before repeating the entire process from the beginning. 

The File Transfer category contains SFTP, RSYNC, and SCP traffic. To emulate users performing file transfers, we created scripts that randomly choose a file out of a list of 15 files of sizes ranging from 1 KB to 1 GB. Then, they randomly decide whether to send the chosen file to or from a remote server. After executing the file transfer, the scripts wait 60 seconds before selecting a new file.

\section{Methods}\label{sec:methods}
In this section, we discuss the features that we used as model input, the specifics of our model architecture, and how we define out of distribution scores.

\subsection{Data Pre-processing}
We group packets into connections based on the five-tuple consisting of the source IP address, destination IP address, source port, destination port, and protocol using the dpkt\footnote{https://github.com/kbandla/dpkt} Python module. We split the connections into 40.96-second intervals\footnote{ A discussion of our selection of 40.96-second intervals and 0.01-second bins is provided in Appendix \ref{appendix:waveletfeatures}.} and discard any intervals containing fewer than 20 packets. For each 40.96-second interval, we obtain the packet timestamps, payload sizes, and directions discretized into 0.01-second bins.  After this preprocessing, the data collection contains 1675 Command \& Control examples, 10498 Chat, 851 File Transfer, 1827 Streaming, and 243 VoIP examples.

\subsection{Feature Construction}
For each 40.96-second interval, we compute aggregate statistics (such as total, average, maximum, etc.) for flow features such as interarrival times and time spent as active or idle.  We compute statistics both for the flow's forward and backward direction. We note here that in our current implementation, ``forward'' and ``backward'' are arbitrary; the first observed packet in the connection is designated as ``forward'' for the remainder of the connection. The full set of aggregate statistics features are listed in Table \ref{tab:tls_features} and are very similar to those in \cite{draper2016characterization}.

\begin{table}[!t]
    \caption{Flow statistic features calculated over each time window.}
    \label{tab:tls_features}
    \centering
    {
        \begin{tabular}{ll}
            \toprule
            \bf{Feature} & \bf{Description} \\ \midrule
            total\_forward & Number of forward packets \\  
            total\_backward & Number of backward packets \\  
            bytes\_per\_sec & Total bytes per second \\  
            total\_bytes\_forward  & Total forward bytes \\  
            total\_bytes\_backward & Total backward bytes \\  
            FIAT (mean, min, max, std) & Forward inter-arrival time \\  
            BIAT (mean, min, max, std) & Backward inter-arrival time \\  
            FlowIAT (mean, min, max, std) & Flow inter-arrival time \\  
            Active (mean, min, max, std) & Total seconds flow is active \\  
            Idle (mean, min, max, std) & Total seconds flow is idle \\ 
            \bottomrule
        \end{tabular}
    }
\end{table}

We also generate wavelet-based features from the 40.96-second intervals using the discrete wavelet transform as done by \cite{shi2017wavelet}. We select the Haar basis as the mother wavelet because it produces the maximum energy to Shannon entropy ratio. Before calculating the wavelet-based features, we extract the detail coefficients from frequency bands 0 to 12 of the wavelet-transformed connection sizes in the forward and backward directions using the PyWavelets Python package \cite{lee2019pywavelets}. Then, we use the detail coefficients of each band to calculate the wavelet-based features from the connection sizes in the forward and backward directions. The wavelet-based features include the relative wavelet energy, Shannon entropy, and absolute means and standard deviations of the detail coefficients as shown in Table \ref{tab:wavelet_features}. The absolute means and standard deviations of the detail coefficients are log-transformed to induce symmetry because the original distributions of these features are heavily right-skewed. See Appendix \ref{appendix:waveletfeatures} for details on how to calculate the wavelet-based features. 

\begin{table}[!b]
    \caption{Wavelet-based features calculated over each time window.}
    \label{tab:wavelet_features}
    \centering
    {
        \begin{tabular}{ll}
            \toprule
            \bf{Feature} & \bf{Description} \\
            \midrule
            RelEng\_Forward & Forward relative wavelet energy \\  
            RelEng\_Backward & Backward relative wavelet energy \\  
            Entropy\_Forward & Forward Shannon entropy \\  
            Entropy\_Backward & Backward Shannon entropy \\  
            MeanDetail\_Forward & Absolute mean of forward detail coefficients \\  
            MeanDetail\_Backward & Absolute mean of backward detail coefficients \\  
            StdDetail\_Forward & Std. deviation of forward detail coefficients \\  
            StdDetail\_Backward & Std. deviation of backward detail coefficients \\  
            \bottomrule
        \end{tabular}
    }
\end{table}

In total, this data pipeline collapses the (up to) gigabytes of network traffic that could flow across 40.96 seconds into two vectors of length 4096 that are then reduced to 129 features per time-window. Some of these features are also correlated and could be winnowed (data not shown) to further sparsen the signal.

\subsection{Classification Model}

Prototypical Networks are neural networks that compute class probabilities using the distances to a set of so-called prototypical examples \cite{snell2017prototypical}, rather than by applying a softmax activation function to the output layer.  The neural network that forms the base of the prototypical network is usually referred to as an embedding function.  We denote the embedding function as $f_{\theta}: \mathbb{R}^D \rightarrow \mathbb{R}^E$, where $D$ is the dimension of the feature space, $E$ is the dimension of the embedding space, and $\theta$ are weights to be learned.  We assume access to a set $X = \{x_i\}$ of examples with corresponding labels $y_i \in \{1, 2, \hdots, K\}$.  We will use the notation $X_{train}$, $X_{cal}$, and $X_{test}$ to refer to partitions of the dataset.  The prototype for each class $k, \ 1 \leq k \leq K$, which we denote as $c_k$ is simply the mean of the embedding of a few $(S)$ examples  of the class (known as the support examples); that is,  
$
c_k = \frac{1}{S}\sum_{s=1}^{S}{f_{\theta}(x_{s}^k)}.
$  
Here, the superscript $k$ indicates the examples come from class $k$.

For a given observation, the class probabilities are calculated by applying the softmax activation function to the distances between the class prototypes and the embedding of the observation. Specifically, the probability of class $k$ for input example $x_i$ is 
\begin{equation}
P(y_i=k | x_i) = \frac{\exp(-d_k(f_{\theta}(x_i), c_k))}{\sum_{k'}{\exp(-d_{k'}(f_{\theta}(x_i), c_{k'}))}},
\label{eq:softmax}
\end{equation}
where each $d_k$ is a distance function on the embedding space.

In our application, we use an embedding model with four fully connected hidden layers each containing 64 neurons with Relu activations.  We perform 25\% dropout on the weights connecting the second and third and the third and fourth layers.  

We learn our model's parameters using the episodic training paradigm \cite{snell2017prototypical}.  In episodic training, for each of $N_{episode}$ episodes, we sample $N_{query}$ query examples and $S_{train}$ support examples.  The prototypes are computed from the support examples and we compute the cross-entropy loss over the query examples.  We summarize our training procedure in Algorithm \ref{alg:train}.  

Throughout the paper, we set $N_{episode}$ to $20000$, $N_{query}$ to 512, and $S_{train}$ to $5$. As in \cite{snell2017prototypical}, the choice of squared Euclidean distance during training roughly corresponds to spherical Gaussian densities in $\mathbb{R}^E$; this motivates our use of the Mahalanobis distance for OOD scoring. 

\begin{algorithm}[!b]
\caption{Train}
\label{alg:train}
\begin{algorithmic}
\Require $X_{train}$, $y_{train}$, $N_{episodes}$, $N_{query}$, $S_{train}$, $f_{\theta}$
\For{$ 1 \leq b \leq N_{episodes}$}
\State Sample support $\{x_s^k\}_{s=1}^{S_{train}}$ for each class $k$ and query examples $\{x_i\}_{i=1}^{N_{query}}$ from $X_{train}$
\State Compute embeddings $f_{\theta}(x_i)$ and $f_{\theta}(x_s^k)$
\State Compute prototypes $c_k = \frac{1}{S_{train}}\sum_{i=s}^{S_{train}}{f_{\theta}(x_{s}^k)}.$
\State Calculate class probabilities as in Equation \eqref{eq:softmax} where each $d_k$ is the Euclidean distance
\State Compute the gradient of the loss over the query examples; update parameters $\theta$
\EndFor
\Ensure $f_{\theta}$
\end{algorithmic}
\end{algorithm}

\subsection{Out of Distribution Scores}
\label{sec:OOD}

We now describe how we obtain OOD uncertainty measures from the prototypical network.  Our OOD score is based on Mahalanobis distance.  Mahalanobis distance is used to measure the distance between a point and a distribution. Given a point $x$ and a distribution with mean $\mu$ and covariance matrix $\Sigma$, the Mahalanobis distance between $x$ and the distribution is
\begin{equation}
M(x; \,\mu, \Sigma) = \sqrt{(x-\mu)^T\Sigma^{\dagger}(x-\mu)}.
\end{equation}
Above, $\Sigma^{\dagger}$ denotes the pseudoinverse of $\Sigma$.  Using the pseudoinverse rather than the matrix inverse handles the case in which the covariance matrix is singular.  For a dataset consisting of $K$ classes, the relative Mahalanobis distance to class $k$ is defined to be the Mahalanobis distance to class $k$ minus the Mahalanobis distance to the overall data distribution \cite{Ren2021}.  That is,
\begin{equation}
M_{rel}^k(x; \, \mu_k, \mu_0, \Sigma, \Sigma_0) = M(x; \, \mu_k, \Sigma) - M(x; \, \mu_0, \Sigma_0),
\end{equation}
where $\mu_k$ is the mean of class $k$, $\mu_0$ is the mean of the dataset as a whole, $\Sigma$ is the covariance matrix of each class $k$, and $\Sigma_0$ is the covariance matrix of the dataset as a whole.  To be clear, $\mu_k = \frac{1}{K}\sum_{s=1}^S{f_{\theta}(x_s^k)}$ and $\mu_0 = \frac{1}{KS}\sum_{k=1}^K \sum_{s=1}^S f_{\theta}(x_s^k)$, while
\begin{equation}
\Sigma = \frac{1}{KS}\sum_{k=1}^K{\sum_{s=1}^S{(f_{\theta}(x_s^k)-\mu_k)(f_{\theta}(x_s^k)-\mu_k)^T}}\, ,
\end{equation}
and
\begin{equation}
\Sigma_0 = \frac{1}{KS}\sum_{k=1}^K{\sum_{s=1}^S{(f_{\theta}(x_s^k)-\mu_0)(f_{\theta}(x_s^k)-\mu_0)^T}}.
\end{equation}

The authors in \cite{Ren2021} argue that subtracting out $M(x; \, \mu_0, \Sigma_0)$ removes the outlier score contribution along dimensions that are not discriminative between in and out of distribution examples.  This is especially important for scenarios in which a large number of dimensions are not discriminative.

After training is complete, we use a support set of maximum size $S_{cal}=100$ sampled from the training data (or all the training data if there are fewer than 100 examples) to fit the means and full covariance matrices used in the relative Mahalanobis distance.  While the relative Mahalanobis distance allows us to rank the examples in the test set according to how outlying they are, it does not immediately suggest a threshold.  To convert the raw distance into an interpretable score, we fit a univariate kernel density estimate to the relative Mahalanobis distances (for each class) computed on a set of held-out in distribution examples $X_{cal}$. The density estimate will be used at test time to define an outlier score between 0 and 1 by integrating under the density to determine the proportion of calibration examples that are more outlying than the test point.  We refer to this process as calibration and summarize our procedure in Algorithm \ref{alg:calibrate}. 

At test time, to compute a p-value for the null hypothesis that a test sample is in-distribution, we calculate the area under the density corresponding to the predicted class for distances larger than the relative Mahalanobis distance of the test point.  To maintain the convention that large scores indicate OOD examples, we take one minus the p-value as our OOD score.  We summarize this procedure in Algorithm \ref{alg:test}.  The computational overhead of the OOD approach is small.  Computing the covariance matrices on the calibration data takes $O(E^2)$, where $E$ is the dimension of the embedding space.  Computing the relative Mahalanobis distances is also $O(E^2)$.  Finally, computing each p-value from the kernel density estimate involves $|X_{cal}|$ evaluations of the normal cumulative distribution function.

\begin{algorithm}
\caption{Calibrate}
\label{alg:calibrate}
\begin{algorithmic}
\Require $X_{train}$, $y_{train}$, $X_{cal}$, $y_{cal}$, $S_{cal}$, $f_{\theta}$
\State Sample support $\{x_s^k\}_{s=1}^{S_{cal}}$ from the training data for each class $k$
\State Compute $\mu_k$, for each class $k$, $\mu_0$, $\Sigma$ and $\Sigma_0$ as defined in Section \ref{sec:OOD} with $S=S_{cal}$
\For {$1 \leq k \leq K$}
\State Compute $\Sigma_k = \frac{1}{S_{cal}}\sum_{s=1}^{S_{cal}}{(f_{\theta}(x_s^k)-\mu_k)(f_{\theta}(x_s^k)-\mu_k)^T}$ (used only in testing phase)
\State Compute relative Mahalanobis distance for calibration examples $M_{rel}^k(f_{\theta}(x_i^k))$
\State Fit Gaussian kernel density estimate $G_k$ to the set of relative Mahalanobis distances $ M_{rel}^k(f_{\theta}(x_i^k)) $
\EndFor
\Ensure $G_k$, $\mu_k$, $\mu_0$, $\Sigma$, $\Sigma_0$, $\Sigma_k$
\end{algorithmic}
\end{algorithm}

\begin{algorithm}
\caption{Test}
\label{alg:test}
\begin{algorithmic}
\Require $G_k$, $\mu_k$, $\mu_0$, $\Sigma$, $\Sigma_0$, $\Sigma_k$ $x_{test}$, $f_{\theta}$
\State Compute predicted class $k^*$ as in Equation \ref{eq:softmax} where $d_k=M(x; \, \mu_k, \mbox{diag}(\Sigma_k))$
\State Compute $r=M_{rel}^{k^*}(f_{\theta}(x_{test}))$
\State Compute p-value $p=\int_r^\infty G_{k^*}(r')dr'$
\Ensure $1-p$
\end{algorithmic}
\end{algorithm}

\section{Experiments and Results}\label{sec:results}
In order to validate our framework, we perform a series of increasingly challenging tests for our machine learning framework. 

\subsection{Model Performance on the Dataset}

\begin{figure}[!t]
    \centering
    {\includegraphics[width=3.3in]{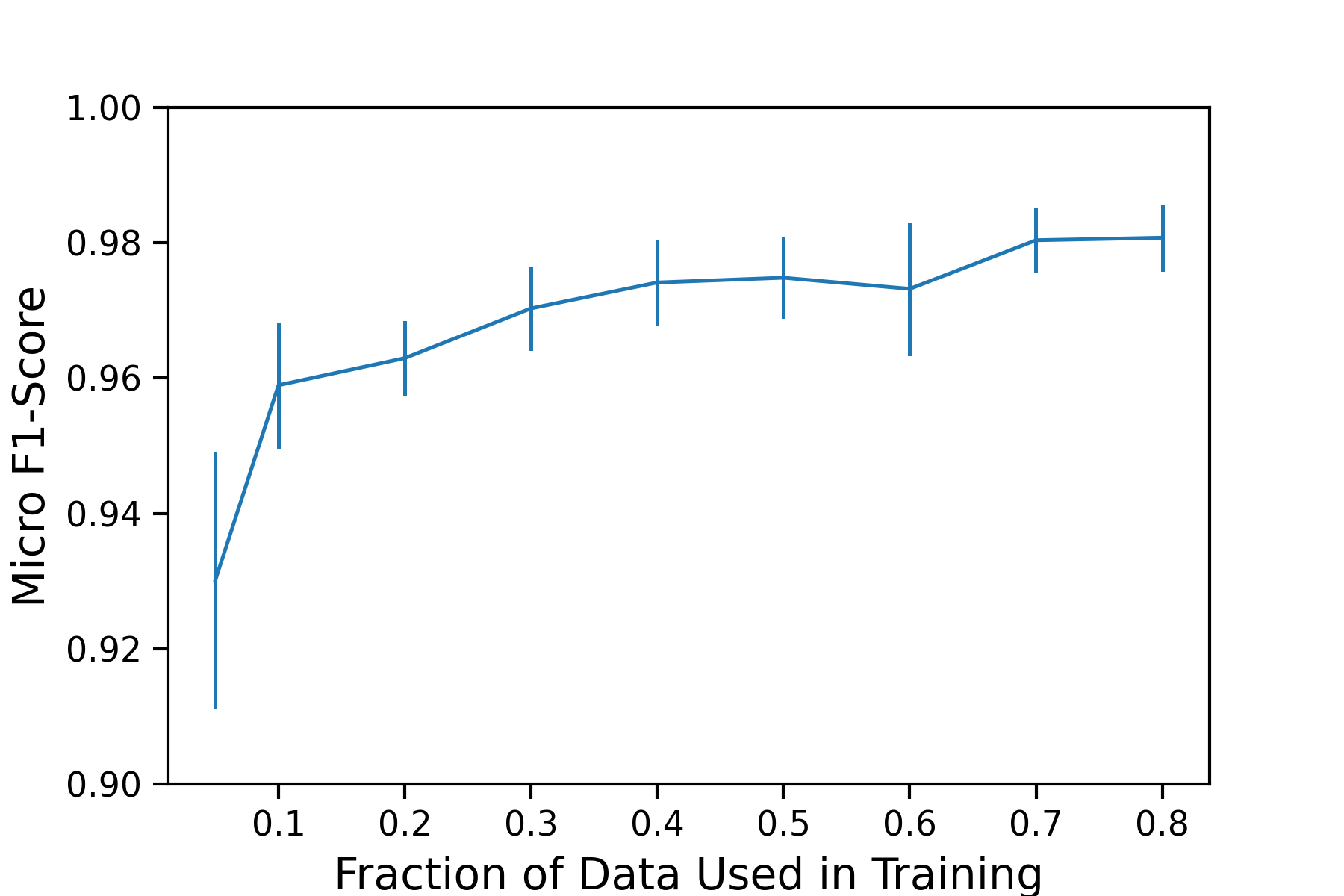}}
    \caption[]{Micro F1-score of the model against the fraction of data used to train the model.  We plot the average F1-score over ten trials.  The error bars indicate one standard deviation.}
    \label{fig:F1}
\end{figure}

\begin{figure}[!t]
    \centering
    {\includegraphics[width=3.3in]{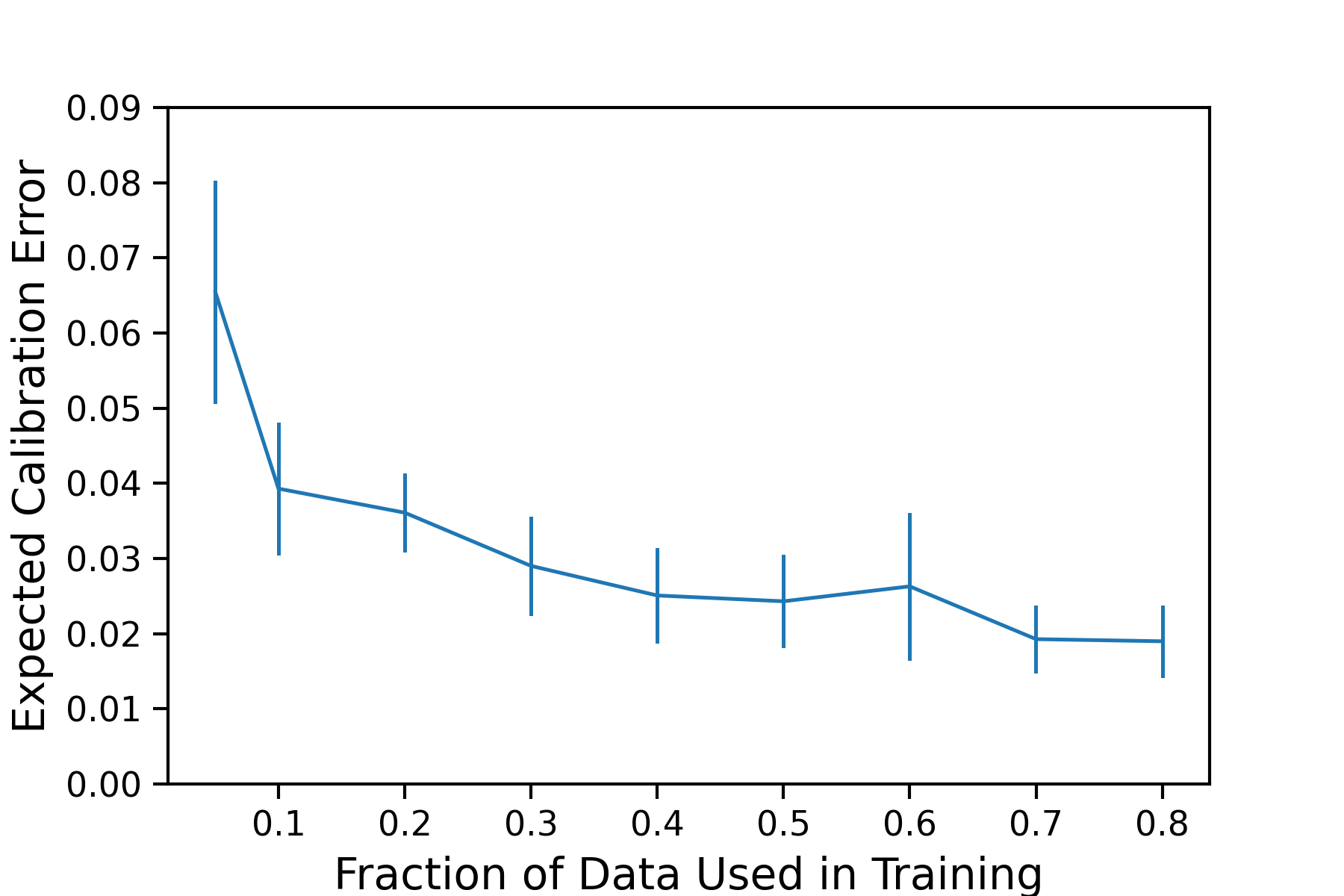}}
    \caption[]{ECE of the model against the fraction of data used to train the model.  We plot the average F1-score over ten trials.  The error bars indicate one standard deviation.}
    \label{fig:ECE}
\end{figure}

First, we train the model on the provided dataset using an increasing, stratified fraction of the dataset and compute the F1-scores (harmonic average of precision and recall) on a held-out test set (20\% of the data).  We also compute the model's Expected Calibration Error (ECE) as a way to measure the quality of its probabilistic forecasts.  ECE measures the discrepancy between a model's predicted probability and the observed accuracy of those predictions.  For more details, see \cite{Guo2017}.  For each training dataset size considered, we repeat the experiment ten times.  The results for F1-score and ECE are shown in Figs.~\ref{fig:F1} and \ref{fig:ECE}.  Clearly, the model's performance improves as we increase the fraction of data provided, both for F1-score and ECE.  It is interesting to note that we achieve quite good performance with only a limited number of training examples.  To call out a specific example, using 10\%\footnote{Technically, only 8\% of the data is used for training, as 20\% of the 10\% is reserved internally by our implementation for generating the p-value distributions.} of the training data corresponds to 118 examples of Command \& Control, 945 examples of Chat, 66 examples of File Transfer, 143 examples of Streaming, and just 19 examples of VoIP.  Even with only this very small training set, we are able to obtain a micro F1-score of nearly 0.96.  Additionally, the calibration error using 10\% of the data for training is about 0.04, which indicates the model's confidence is fairly consistent with its accuracy. Marginal gains in calibration (lower ECE values) can be achieved via neural network ensembles as in \cite{Brown2020} (data not shown); we speculate that the relatively shallow network, dropout, and Prototypical Network architecture helped to enable the low ECE values.  
Fig.~\ref{fig:normal_cm} depicts the average confusion matrix for each of the ten training runs using an 80/20 train/test split, indicating that the model performs well in every category, the worst confusion being 3\% of File Transfer examples being incorrectly predicted as VoIP.

\begin{figure}[htbp]
    \centering
   \def\myConfMat{{
{0.97, 0.00, 0.00, 0.00, 0.02},
{0.00, 0.99, 0.00, 0.00, 0.00},
{0.01, 0.01, 0.94, 0.01, 0.03},
{0.00, 0.01, 0.01, 0.95, 0.03},
{0.00, 0.00, 0.00, 0.00, 1.00}
}}

\def\classNames{{"C2","CHAT","FT","STREAM","VOIP"}}

\def\numClasses{5}

\def\myScale{1.1} 
\begin{tikzpicture}[
    scale = \myScale,
    font={\footnotesize}, 
    ]

\tikzset{vertical label/.style={rotate=90,anchor=east}} 
\tikzset{diagonal label/.style={rotate=45,anchor=north east}}

\foreach \y in {1,...,\numClasses}
{
    \node [anchor=east] at (0.4,-\y) {\pgfmathparse{\classNames[\y-1]}\pgfmathresult}; 
    
    \foreach \x in {1,...,\numClasses}  
    {
    \def\totSamples{0.0}
    \foreach \ll in {1,...,\numClasses}
    {
        \pgfmathparse{\myConfMat[\ll-1][\x-1]}   
        \xdef\totSamples{\totSamples+\pgfmathresult} 
    }
    \pgfmathparse{\totSamples} \xdef\totSamples{\pgfmathresult}  
    
    \begin{scope}[shift={(\x,-\y)}]
        \def\mVal{\myConfMat[\y-1][\x-1]} 
        \pgfmathsetmacro{\r}{\mVal}   %
        \pgfmathtruncatemacro{\p}{round(\r*100)}
        \pgfmathtruncatemacro{\pone}{100-\p}
        \coordinate (C) at (0,0);
        \ifthenelse{\p<50}{\def\txtcol{black}}{\def\txtcol{white}} 
        \node[
            draw=black,                 
            text=\txtcol,         
            align=center,         
            fill=blue!\p,        
            minimum size=\myScale*10mm,    
            inner sep=0,          
            ] (C) {\r};     
        \ifthenelse{\y=\numClasses}{
        \node [diagonal label] at ($(C)-(0,0.55)$) 
        {\pgfmathparse{\classNames[\x-1]}\pgfmathresult};}{}
    \end{scope}
    }
}
\coordinate (yaxis) at (-0.3,0.0-\numClasses/2);  
\coordinate (xaxis) at (0.2+\numClasses/2, -\numClasses-1.35); 
\node [vertical label] at (yaxis) {Actual};
\node []               at (xaxis) {Predicted};
\end{tikzpicture} 
    \caption{Average confusion matrix for 10 runs using a randomized 80/20 train/test split (FT indicates File Transfer).}
    \label{fig:normal_cm}
\end{figure}

\subsection{Model Performance on UTMobileNet2021}
In order to evaluate our approach on a greater variety of labels, we evaluated our model on the UTMobileNet2021 encrypted traffic dataset, which contains data from 17 unique traffic applications. Using the above settings, we found that our approach retained 80\% accuracy across all applications using an 80/20 train/test split, which is on par with their best reported results \cite{heng_utmobilenettraffic2021_2021}. We also tested the effect of adding multiple labels to non-overlapping applications from our own dataset, and we found that for 5, 10, 15, and 20 application labels, our approach had overall accuracies of 93\%, 87\%, 80\%, and 80\%, indicating that our per-application performance degrades a bit but remains competitive. For more details, see Table \ref{t:utmobilenet} in the Appendix.

\subsection{Model Performance in a New Environment}

Having established the ``usual'' ML metrics, we then turn to a more challenging problem of validation outside of the training dataset. To do these experiments, we used Wireshark to capture network traffic on a live, enterprise network, and then tested our model on this new traffic.  We consider three increasingly challenging scenarios.  First, can the model recognize traffic from an existing application in an existing category?  Second, can the model recognize new applications in an existing category?  Finally, can the model detect data that is out of distribution (comes from a new category)?

To answer the first question, we collected 40 additional instances of YouTube (an application in the Streaming category).  We trained the model on 80\% of our training dataset and then tested on the YouTube data.  We repeated the experiment ten times.  The results were quite good.  For the YouTube traffic, the model predicted the correct class 40 out of 40 times in nine of the trials and 39 out of 40 times in the other trial, only predicting one of the examples to have an OOD score above 0.95 on average\footnote{We have selected OOD scores above 0.95 as a threshold of interest, as it corresponds to a p-value of 0.05 or lower for rejecting the null hypothesis that the test samples are in-distribution.}. Thus, even though the YouTube traffic was collected on a different network than the one that the training data came from, the model was still able to recognize its class (Streaming). This is an encouraging result, although we do not rely on this transferability holding true for all applications in the training set, as will be demonstrated next.

To answer the second question, we collected data from three applications that do not appear in the training data (Avaya, Samba (SMB), and the Code42 automated file backup program).  In our labeling schema, we consider Avaya to be a VoIP application and SMB and Code42 to be File Transfer.  We collected 123 examples of Avaya, 141 examples of SMB, and 202 examples of Code42.  As before, we trained on 80\% of our training dataset (for ten trials) and tested on the new applications.  Avaya was easily recognized by the model as VoIP traffic, correctly labeling all 123 examples in each trial, although the model was less certain about them, assigning about half of them OOD scores above 0.95. SMB and Code42, however, were not initially recognized as File Transfer.  Encouragingly, the model tended to give these examples a high OOD score.  Specifically, across the ten trials, an average of 78\% of the SMB traffic received an OOD score above 0.95, and an average of about 63\% of the file backup program received an OOD score above 0.95.

\begin{figure}[!b]
    \centering
    {\includegraphics[width=3.3in]{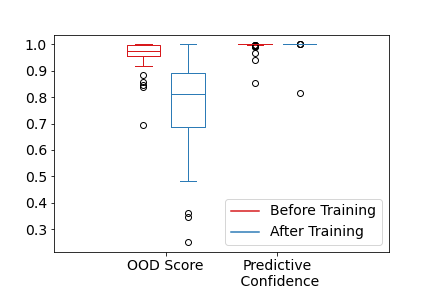}}
    \caption[]{Model uncertainties for SMB from an enterprise network before (using a pretrained model from our dataset) and after retraining (adding 91 labeled examples).}
    \label{fig:smb_ood}
\end{figure}

To see whether the model could learn to classify SMB and Code42 as File Transfer, given a few training examples, we investigate retraining the model.  For the case of SMB, we add 91 examples to our training set and test on the remaining 50.  After retraining, the model is able to correctly classify an average of 92\% of the test samples.  In addition, the OOD scores are much lower.  After retraining, only an average of 10\% of examples had OOD scores above 0.95 (left side of Figure \ref{fig:smb_ood}).  Figure \ref{fig:smb_ood} also provides the predictive confidences before and after retraining with examples of SMB, which refers to the probability associated with the class prediction (the highest probability class among known classes). The distributions exhibit high confidences both before and after retraining.
This indicates that it is important to not rely solely on the predictive confidences for uncertainty quantification, as they are essentially meaningless when applied to OOD test examples.

\begin{figure}[!b]
    \centering
    {\includegraphics[width=3.3in]{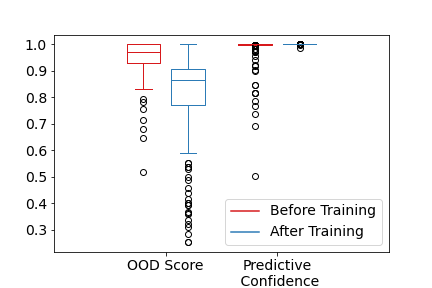}}
    \caption[]{Model uncertainties for Code42 from an enterprise network before (using a pretrained model from our dataset) and after retraining (adding 42 labeled examples).}
    \label{fig:file_backup}
\end{figure}

In the case of Code42, we add 42 examples to our training set and test on the remaining 160.  The model now correctly classifies an average of 98\% of the test samples.  As with SMB, the OOD scores are now much lower as well, with only an average of 5\% with scores above 0.95.  In Figure \ref{fig:file_backup}, we show OOD score and predictive confidence before and after training with examples of Code42.

\begin{figure}[!t]
    \centering
    {\includegraphics[width=3.3in]{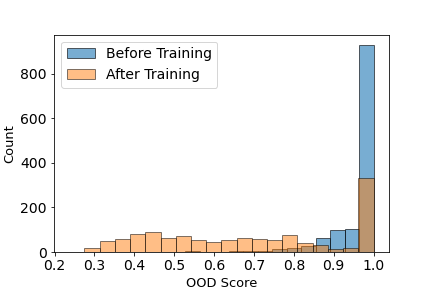}}
    \caption[]{Distribution of Zoom OOD scores on a held-out validation set of 128 examples (sum of 10 runs) using the pretrained model from our dataset (``Before Training'') and after retraining with 195 labeled examples (``After Training'').}
    \label{fig:zoom_hist}
\end{figure}

To answer the final question, we collected 323 examples of Zoom, which is a new application of a new category one could label as Video TeleConferencing (VTC).  Since VTC is a new class, we expect that the model will give high OOD scores to the Zoom examples.  This ended up being the case, as an average of 78\% of examples had an OOD score above 0.95.  We then retrained our model with 195 Zoom examples added to the training set (with the new VTC label).  
After retraining, the fraction of OOD scores above 0.95 is reduced to 26\%.  Figure~\ref{fig:zoom_hist} depicts the distribution of OOD scores before and after training on a held-out validation set of 128 Zoom examples; the post-training results are more uniform, as expected.
Across 10 trials, the model correctly labeled an average of 73\% of the validation examples as Zoom, generally confusing the errors as VoIP. We note that this accuracy is lower than the previous examples, however, about 97\% of the incorrectly labeled examples still have OOD scores above 0.95, indicating that these errors can easily be screened out due to their model uncertainty. It perhaps indicates that Zoom traffic behaves differently depending upon its mode of usage (screen sharing, video, or audio-only); it may be possible to cluster self-similar OOD examples in the embedding space to assist analysts with new labeling.

\subsection{Model Performance on Padded Packets}

In order to address any concerns about the model's resilience to encryption protocols that pad packets to be of uniform size, we repeated the initial testing on the collected dataset but masked the sizes of all packets to be 1500 bytes before feature generation. Upon retraining ten times with 80/20 train/test splits, the model retained an average F1-score of $0.971\pm 0.008$ on the dataset. This is a marginal drop from the raw, unpadded data, at $0.982\pm0.004$.  These results indicate that our model is not much affected by the effective removal of packet size information.  In addition, this result presents initial evidence that packet padding alone may not offer substantial privacy benefits, at least at the level of granularity of traffic categories.

\subsection{Baseline OOD Performance}

\begin{figure}[!t]
    \centering
    {\includegraphics[width=3.3in]{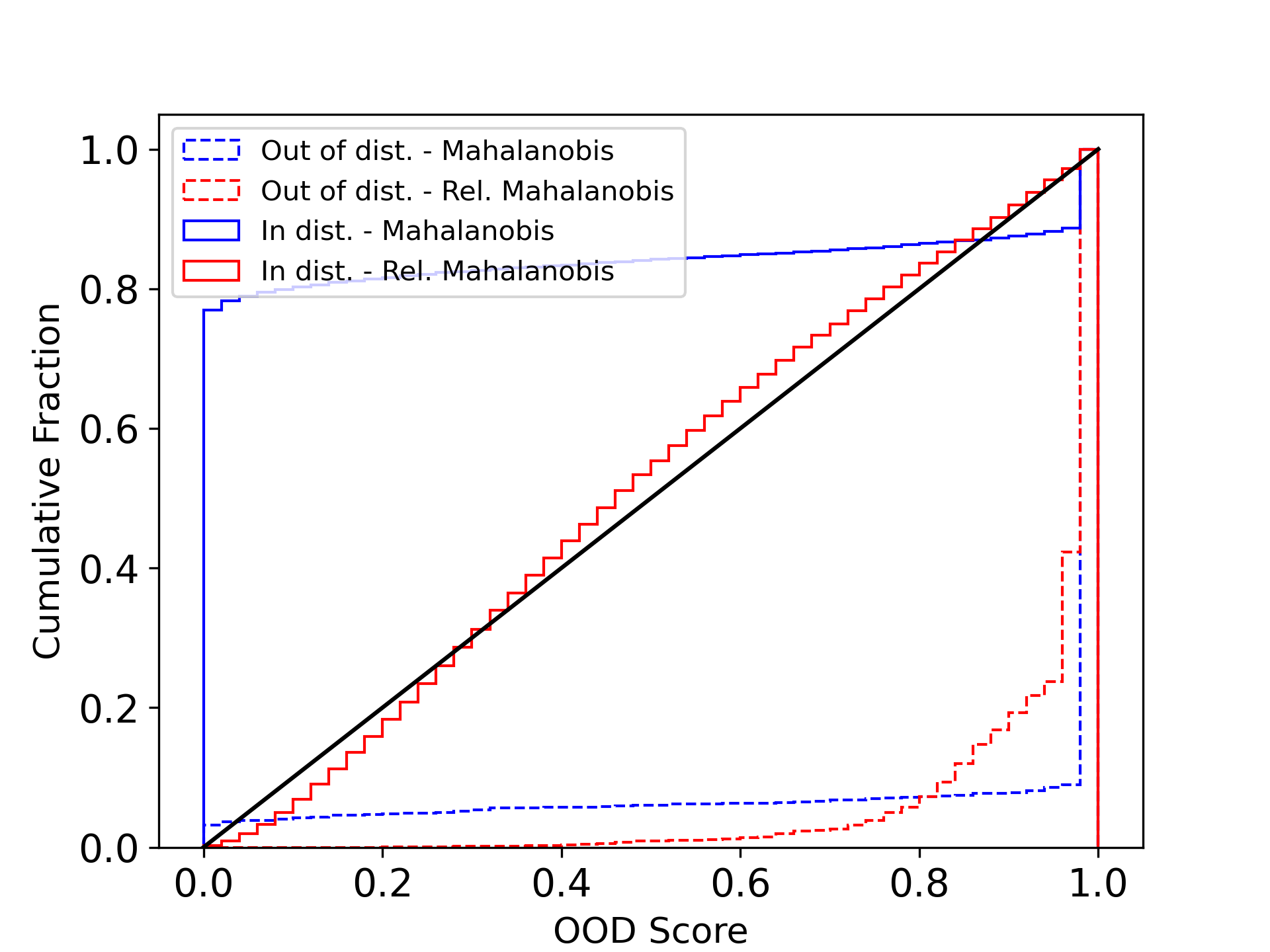}}
    \caption[]{Empirical cumulative distribution function of OOD scores on a held-out set of 116 Zoom examples (out of distribution data) and a held-out set of in distribution data.  OOD scores are produced using the squared Mahalanobis distance and $\chi^2$ distribution as well as the Relative Mahalanobis distance and a KDE fit on calibration examples.}
    \label{fig:zoom_CDF}
\end{figure}

Finally, we compare our OOD approach to a baseline approach. In our baseline, rather than fitting a KDE on a calibration set to compute an outlier score, we note that the squared Mahalanobis distance follows a $\chi^2$ distribution with 64 (the dimension of our embedding space) degrees of freedom.  We use this distribution to compute p-values for new observations.  As before, our OOD score is simply one minus the p-value.  This assumes, of course, that the embeddings are actually normally distributed according to the model in Section \ref{sec:methods}.   In Figure \ref{fig:zoom_CDF}, we evaluate our model on a held-out test set of in distribution data as well as 116 examples of Zoom traffic (which is out of distribution).  

We first examine the in distribution data.  Notice that the OOD scores based on the squared Mahalanobis distance and $\chi^2$ distribution tend to be either close to 0 or close to 1.  This stands in contrast to the scores produced by the relative Mahalanobis distance and the KDE, which are roughly uniform.  Thus, if we set the OOD detection threshold to $\alpha$, then we expect to filter out $1-\alpha$-percent of the in distribution data, when using the approach proposed in this paper.  

For out of distribution data, both approaches tend to assign highs scores, for the most part.  The squared Mahalanobis distance-based score again tends to be either very close to 1 or very close to 0, while the relative Mahalanobis-based score varies more smoothly.

\section{Discussion}\label{sec:discussion}
The results of the previous section appear very encouraging, particularly since (1) the framework performs effectively, achieving high accuracy without requiring rigorous feature optimizations and on only timing and size information, (2) the dataset and model extended in part to another network for YouTube and VoIP, and (3) the model ``degraded gracefully'' when confronted with novel application data and was quick to retrain on limited data. We envision this framework to be used in a tool that would allow network analysts to systematically build insight for applications of interest in conjunction with other analysis techniques (malware signatures, website fingerprinting, etc.) to supplement where they fail. 

Shorter window-lengths could hypothetically provide more label opportunities over the entire observation period, but arbitrarily short window lengths would reduce accuracy (additional analysis is provided in Figure \ref{fig:wavelet_study}). Longer window-lengths within a VPN would also be more likely to merge multiple application types into the same sample, and our approach only predicts one application per time window. We also note that this framework does not leverage continuity between time-windows as of yet, and certain connections can tend to be shorter (loading a web page) or longer (streaming movies). 

Our approach can also be attacked and degraded by manipulating the traffic, e.g., by adding ``cover packets'' to flood connections or mask timings, re-encoding traffic through another protocol \cite{sharma_camoufler_2021} or by leveraging adversarial ML \cite{nasr_defeating_2021, Geneva}. However, our approach still retains an advantage in that it forces an adversary to work harder to obfuscate the nature of their connections. We also note that our approach could be used to ``red-team'' censorship evasion development by incorporating uncertainty in addition to the standard accuracy metrics. 

We finally comment that although our dataset does not comprehensively capture the diversity of Internet applications, it was enough to ``generate intuition'' for a generic ML framework to predict time-windowed, encrypted application traffic on a different network, and we have demonstrated the importance of treating uncertainty as a first-class feature to enable a systematic extension of the dataset onto other networks or application categories of interest.

\section{Conclusions and Future Work}\label{sec:conclusions}
We presented a labeled dataset of VPN-encrypted and unencrypted network traffic to facilitate machine learning applications seeking to leverage packet size and timing information to predict the traffic's most likely application categories. We discussed potential issues and possible workarounds for a popular alternative dataset, which motivated the organization of our dataset. We showed a machine learning framework leveraging wavelet-based and statistical features on discrete, sampled time windows in a Prototypical Network architecture that achieved high F1 scores (0.98) on the labeled dataset. Further, we showed the architecture could be trained on a small, random fraction of the dataset, achieving an accuracy of 0.96 on VOIP using fewer than 14 minutes of data. Additionally, we demonstrated low calibration errors for predictive confidence (3\%). We also tested the model on application traffic (YouTube and a new VoIP application) from an enterprise network and found it extended accurately. 

We used a technique for constructing an out-of-distribution (OOD) score based on p-values of the relative Mahalanobis distance on in-distribution data to reliably indicate model degradation in a series of increasingly challenging experiments. In all cases, the model retrained quickly with relatively few examples to learn to incorporate the new applications into a known category or assign them to a new one. We additionally demonstrated that the framework retains very high performance (F1-score of 0.97) even when all packet sizes are uniform, such as occurs in encryption protocols that ensure all packets are padded to the same size before encryption.

Future work should incorporate the identification of concurrent applications inside a VPN (multi-label predictions), as our approach currently only predicts one class per time interval. Another interesting avenue of research would be to look for structure in OOD examples, perhaps via clustering, and suggest possible new classes to the user. Finally, future work should consider adversarial methods that attack the framework, such as flooding connections with dummy packets, re-encoding traffic through other protocols, or adversarial ML techniques that target both the accuracy and the uncertainty scores.

\section*{Acknowledgment}
 DISTRIBUTION STATEMENT A. Approved for public release. Distribution is unlimited.
This material is based upon work supported by USCYBERCOM under Air Force Contract No. FA8702-15-D-0001. Any opinions, findings, conclusions or recommendations expressed in this material are those of the author(s) and do not necessarily reflect the views of USCYBERCOM.

\appendix
\section*{Filename-to-Category Mappings}
\label{appendix:filemappings}
We use the mappings shown in Table \ref{tab:vast_files} to assign class labels to each PCAP file. If any of the words in the ``File Keyword'' column occurs in the column name, then we assign the label in the left column. 
\begin{table}[htbp]
\caption{Filename-to-category labeling scheme.}
\label{tab:vast_files}
\centering
    {
    \begin{tabular}{ll}
    \toprule
    \bf{Traffic Category} & \bf{File Keyword}  \\ \midrule
    STREAMING & vimeo, netflix, youtube \\  
    CHAT & chat \\  
    C2 & ssh, rdp \\
    FILE\_TRANSFER & sftp, rsync, scp \\
    VOIP             & voip \\ \bottomrule
    \end{tabular}
    }
\end{table}

\section*{Wavelet-based Features}
\label{appendix:waveletfeatures}
This section provides details on how to calculate the wavelet-based features extracted from the wavelet-transformed connection sizes in the forward and backward directions for frequency bands 0 to 12. The wavelet-based features include the relative wavelet energy, Shannon entropy, and absolute means and standard deviations of the detail coefficients. 

The continuous wavelet transform coefficients are defined by the equation:
\begin{equation}
\label{eq:cont_wavelet}
    d(a,b) = \frac{1}{\sqrt{|a|}} \int_{0}^{T} x(t) \overline{\psi} \left(\frac{t-b}{a}\right) dt\, ,
\end{equation}
where $x(t)$ is the time-dependent signal over the observation period $[0,T]$ (for instance, bytes transferred in a flow during a sample interval), $\overline{\psi} \left(\frac{t-b}{a}\right)$ is the wavelet function, $a$ is the scale factor, and $b$ is the translation factor. This work employs the discrete wavelet transformation, which restricts $a$ and $b$ to discrete values that balance temporal and frequency resolution while reducing computational burden. We employ the shift-invariant (stationary) wavelet transformation to ensure that the resulting features are insensitive to signal translations, otherwise known as the ``algorithm a-trous'' \cite{holschneider1990real, Percival2000Wavelet}.

The time-dependent signal, $x(t)$, required for the wavelet transform, is represented by a discrete vector over an observation period from $[0,T]$ that is discretized into $N$ smaller time bins of width $\Delta t$. Within each $\Delta t$ time interval, all of the packet statistics are aggregated. For instance, let $t_j, y_j$ be the packet timestamp and size for packet $j$ in a series of $J$ flow-aggregated packets with $t_0=0$. Then a vector version of $x(t)$, appropriate for the wavelet transform, is defined by $x_n(t)$, where $T=N \Delta t$, $n \in [0,N-1]$, and
$$
x_n(t) = \sum_{j=0}^J y_j \chi_{j}(t)\, ; \, 
\chi_j(t) = \begin{cases} 1 &  n \Delta t \le t_j < (n+1)\Delta t\\ 
                          0 & \text{else} \end{cases}.
$$
We enforce that the length of the vector $x_n$ be an integer power of 2 ($N=2^K$ for some $K$) by judiciously selecting $\Delta t$ and $T$, or by reflecting the signal in our pipeline.
Applying Eq.~\ref{eq:cont_wavelet} for fixed values of $a\in 2^k$ and $b\in x_n$ gives a rectangular coefficient matrix of the form $d_{n,k}$, where $n\in [0,N-1]$ indexes the coefficients in time and $k\in[0,K]$ indexes the coefficients in frequency-band. 

\subsection*{Relative Wavelet Energy}
For each frequency band $k$, the relative wavelet energy $\rho_k$ is given by 
\begin{equation}
    \rho_k = \frac{E_k}{E_{total}} \quad \text{for}\ k=0,1,\ldots,K\, ,
\end{equation}
where $E_k$ is the energy of frequency band $k$ 
\begin{equation}
    E_k = \sum_n |d_{n,k}|^2 \quad \text{for}\ n=1,2,\ldots, N\, ,
\end{equation}
and $E_{total}$ is the total energy of the signal after wavelet decomposition
\begin{equation}
    E_{total} = \sum_k E_k \quad \text{for}\ k=0,1,\ldots,K\, .
\end{equation}

\subsection*{Wavelet Shannon Entropy}
For each frequency band $k$, the wavelet Shannon entropy $S_k$ is given by
\begin{equation}
    S_k = -\sum_n p_n \log(p_n) \quad \text{for}\ n=1,2,\ldots,N\, ,
\end{equation}
where $p_n$ is the fraction of energy of the $n$-th detail coefficient,
\begin{equation}
    p_n = \frac{d_{n,k}^2}{E_k}\, .
\end{equation}

\subsection*{Detail Coefficient Statistics}
For each frequency band $k$, the absolute mean of the detail coefficients $\mu_k$ is given by
\begin{equation}
    \mu_k = \frac{1}{N} \sum_n |d_{n,k}| \quad \text{for}\ n=1,2,\dots,N \, ,
\end{equation}
and the standard deviation of the detail coefficients $\sigma_n$ is given by
\begin{equation}
    \sigma_k = \sqrt{\frac{1}{N} \sum_n (\mu_k-d_{n,k})^2} \quad \text{for}\ n=1,2,\ldots, N\, .
\end{equation}

\begin{figure}
    \centering
    {\includegraphics[width=3.4in]{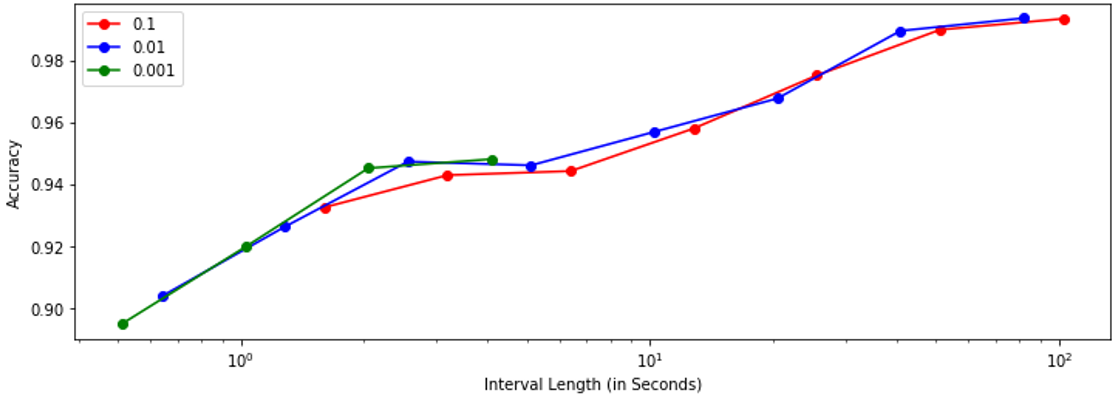}}
    \caption[Wavelet discretization study]{A study in the sensitivity of the overall machine learning accuracy with respect to different choices of window-length $T$ and three time-bin sizes $\Delta t$, 0.1, 0.01, and 0.001 seconds. Note the semi-log x-scale.}
    \label{fig:wavelet_study}
\end{figure}

Figure~\ref{fig:wavelet_study} depicts the results of a range of experiments using three time bins ($\Delta t$ of 1, 10, and 100 milliseconds) and variable time-windows, $T$. The general trend is that, the longer the sample window, $T$, the more accurate the model. However, this has the tradeoff that longer windows require longer vectors, particularly for small time bins (note that we capped the number of wavelet bands at 13 to limit feature-computation requirements). The second-to-rightmost blue datapoint represents $(\Delta t, T)$=$(0.01, 40.96)$, the selected value for the analysis in this paper, as this neared the maximum accuracy and used an acceptable vector length for computing the wavelet features.

\section*{UTMobileDataNet2021 Study}
Table \ref{t:utmobilenet} provides the random classes used to study the effects of adding additional classes on the accuracy of our model.
\begin{table}
\caption{UTMobileNet2021 multiple classes study; bold classes indicate applications supplemented from our dataset.}
\label{t:utmobilenet}
\begin{tabularx}{0.5\textwidth}{lXl}
\toprule
Number of Classes & Applications & Accuracy \\ \midrule
  5-class &  hangout, youtube, skype, twitter, hulu & 93\% \\ \midrule
 10-class &  hangout, instagram, google-maps, dropbox, gmail,
             spotify, twitter, hulu, facebook, reddit & 87\% \\  \midrule
 15-class &  skype, facebook, pandora, spotify, twitter,
             gmail, reddit, messenger, hulu, netflix,  
             hangout, instagram, youtube, google-maps, pinterest & 80\% \\ \midrule
 20-class & dropbox, facebook, gmail, google-drive, google-maps, 
            hangout, hulu, instagram, messenger, netflix, 
            pandora, pinterest, reddit, skype, spotify, 
            twitter, youtube, \textbf{zoiper5, rdp, ssh}  & 80\% \\ \bottomrule
\end{tabularx}
\end{table}
We note that we did not attempt to optimize which applications were chosen for the few-application study, and that the applications selected are random sub-sample of the UTMobileNet2021 Dataset \cite{heng_utmobilenettraffic2021_2021}, with the exception of the 20-class case, which uses 3 classes from our dataset (zoiper5, rdp, ssh). We also note that Heng et al.\'s paper appears to indicate that there are only 16 classes in this dataset, but this excludes their Skype data, which accounts for the ``extra'' label. 

\bibliography{references}
\bibliographystyle{IEEEtran}

\end{document}